# The Hubbard model within the equations of motion approach

F. MANCINI and A. AVELLA

Dipartimento di Fisica 'E.R. Caianiello' – Unità INFM di Salerno
Università degli Studi di Salerno, I-84081 Baronissi (SA), Italy

## Abstract

The Hubbard model plays a special role in condensed matter theory as it is considered to be the simplest Hamiltonian model one can write in order to describe anomalous physical properties of some class of real materials. Unfortunately, this model is not exactly solved except in some limits and therefore one should resort to analytical methods, like the Equations of Motion Approach, or to numerical techniques in order to attain a description of its relevant features in the whole range of physical parameters (interaction, filling and temperature). In this paper, the Composite Operator Method, which exploits the above mentioned analytical technique, is presented and systematically applied in order to get information about the behaviour of all relevant properties of the model (local, thermodynamic, single- and two-particle properties) in comparison with many other analytical techniques, the above cited known limits and numerical simulations. Within this approach, the Hubbard model is also shown to be capable of describing some anomalous behaviour of cuprate superconductors.

## Contents

## Introduction

In the last few decades, the experimental and theoretical study of materials presenting anomalous properties of technological relevance (superconductivity, magneto-resistance, etc.) has raised more and more interest also thanks to the possibilities, just recently opened, to design and synthesize materials at the micro- and nanoscopical level and to characterize them with quite impressive resolutions. Unfortunately, the significant improvements of the experimental techniques do not correspond to significant enlightenment coming from the theoretical approaches. Many of the anomalous properties under analysis are due to the strength and complexity of the correlations present in these systems. According to this, in order to attempt a microscopical description and to gain some comprehension of the physical properties observed in these materials, new methods of theoretical investigation must be formulated.

In the last ten years, we have been developing a theoretical scheme, named the Composite Operator Method (COM), for strongly correlated systems mainly based on two concepts: the use of composite fields to describe the fundamental excitations dynamically generated in a complex system and the use of symmetry constraints to bind the dynamics to the correct Hilbert space. COM has shown to be capable of catching the physics and describing the features of many strongly interacting systems and, in particular, of the Hubbard model and of the materials mimed by it (e.g., the cuprate superconductors). The Hubbard model is recognized as the archetype for the strongly correlated class of Hamiltonian systems: competition of itinerancy and localization, spin and charge ordering, metal–insulator transition, complex phase diagram, etc.

In this paper we wish to accomplish three main goals: first, to give a detailed presentation of the Composite Operator Method; second, to provide a quite extensive review of the physical properties of the Hubbard model in known exact limits (noninteracting, atomic and $1D$), within the two-pole approximation and by means of numerical techniques (quantum Monte Carlo, Lanczos, Exact Diagonalization); third, to summarize the results (regarding local, thermodynamic, single-particle, two-particle and, in general, response properties) of the application of the COM to the Hubbard model within various boundary conditions, to compare them with the numerical simulations and to interpret the experimental data for the cuprate superconductors. The aim is to show the peculiarities and capabilities of this method as a general approach for strongly correlated systems and, in combination with the Hubbard model, its potentialities to describe, on the microscopical level, both relevant theoretical issues and anomalous experimental features of real materials.

The Composite Operator Method tries to give global answers (i.e., in the whole space of model and physical parameters: interaction strength $U$, temperature $T$, filling $n$) in the quest for a deep comprehension of the Hubbard model properties. The model response is profoundly different according to the region of the latter space we decide to explore. Other approximations focus just on one region and usually give wrong results in the rest of the parameter space. We will see that the comparisons with both the numerical simulations and the experimental data show very good agreement on the whole parameter space. According to this, it is fully justified to adopt a very high degree of confidence in the results themselves and in the related microscopical interpretations of the features of the model and materials. Anyway, we are aware that the approximation used in this review, the two-pole approximation, misses relevant features at low temperature/frequencies. Then, it is worth noticing that this drawback is due to the approximation chosen and not to the COM itself. In fact, this latter allows the inclusion of a fully momentum- and frequency-dependent self-energy in the scheme and we are currently working in this direction.